# GHz bandwidth electro-optics of a single self-assembled quantum dot in a charge-tunable device

Jonathan H. Prechtel[1], Paul A. Dalgarno[2], Robert H. Hadfield[3], Jamie McFarlane[3], Antonio Badolato[4], Pierre M. Petroff[5] and Richard J. Warburton[1]

[1]*Department of Physics, University of Basel, Klingelbergstrasse 82, 4056 Basel, Switzerland*
[2]*Scottish Universities Physics Alliance (SUPA) and School of Physics and Astronomy,*
*University of St. Andrews, North Haugh, Fife, KY16 9SS, UK*
[3]*Scottish Universities Physics Alliance (SUPA) and School of Engineering and Physical Sciences,*
*Heriot-Watt University, Edinburgh, EH14 4AS, UK*
[4]*Department of Physics and Astronomy, University of Rochester, Rochester, New York 14627, USA*
[5]*Department of Materials, University of California, Santa Barbara, California 93106, USA*

E-mail: jonathan.prechtel@unibas.ch

The response of a single InGaAs quantum dot, embedded in a miniaturized charge-tunable device, to an applied GHz bandwidth electrical pulse is investigated via its optical response. Quantum dot response times of $1.0 \pm 0.1$ ns are characterized via several different measurement techniques, demonstrating GHz bandwidth electrical control. Furthermore a novel optical detection technique based on resonant electron-hole pair generation in the hybridization region is used to map fully the voltage pulse experienced by the quantum dot, showing in this case a simple exponential rise.

**I. Introduction**

There is currently significant interest in applying high frequency electronics in the GHz range to the development of quantum information technologies. Modern GHz technology, developed originally for wireless communications, is capable of generating electrical pulses with short rise and fall times and with very little amplitude and phase noise, attractive features for quantum control. Furthermore, by utilizing a common technology, quantum and classical information technologies can be easily interfaced. GHz electronics underpins the activity in superconducting qubits [1,2]; coherent manipulation of a spin qubit with GHz electrical control has been demonstrated [3,4]; and recently, these high frequency electrical techniques have also been applied to a trapped ion [5].

Self-assembled quantum dots are potentially a key element in quantum communication systems. A single self-assembled quantum dot is a robust, narrowband and fast source of single photons [6]. A single self-assembled quantum dot can also be used as a spin qubit using either an electron [7,8,9] or hole spin [10,11,12,13], potentially with applications as a quantum repeater or quantum information processor. A key advantage of self-assembled quantum dots is the ability to embed the quantum dots into semiconductor heterostructures, allowing for instance single electron charging in a vertical tunnelling device [14,15]. Another key advantage is the use of post-growth processing, allowing for instance the creation of microcavity structures [16], photonic nanowires [17] and, as is the case here, miniaturized electro-optic devices. Electrical control of self-assembled quantum dots at GHz frequencies for dark-to-bright exciton conversion [18], single photon generation [6] and exciton coherent control [19] has already been demonstrated. A recent breakthrough enabled the entanglement of two spins in a self-assembled quantum dot molecule with electrical pulses [20].

There are two fundamental challenges in this area. The first is the creation of a high bandwidth electrical connection between the room temperature source and the self-assembled quantum dot at low temperature, maintaining optical access and optical alignment. The second is the accurate and reliable monitoring of the actual dot response to an external GHz driving pulse. As shown previously, photolithography defined miniaturized charge-tunable devices with high speed cabling and electronics offer a potential solution to the first challenge [18]. However, characterizing the dot response has remained elusive. Traditional electrical characterization, for instance with a network analyser, is not entirely suitable as it monitors the response of the entire system, in particular the electrical characteristics of the entire macroscopic device at low temperature, rather than the response of the active element, the quantum dot. Instead, it is much better to use the quantum dot itself as a probe of the electrical pulses. In particular, time-correlated-single-photon-counting (TCSPC) techniques [21] can easily achieve sub-100 ps jitter, allowing the optical response to be measured on timescales corresponding to the jitter of the electrical pulse generator.

In this paper we present three complementary optical techniques to map accurately the response of a single self-assembled quantum dot embedded within a miniaturized charge-tunable device architecture. In each case we utilize the spontaneous emission from the quantum dot itself. We report GHz response functions and in particular demonstrate a technique that accurately maps the voltage response of a single quantum dot to an ultrafast electrical pulse, showing in this case a simple mono-exponential rise.

## II. Experimental setup

InGaAs quantum dots were grown within a GaAs charge-tunable heterostructure by molecular beam epitaxy with a density gradient across the wafer [14]. As shown in Fig 1(a), the quantum dots are located 25 nm above a heavily n-doped GaAs back contact (n = $4 \times 10^{18}$ cm$^{-3}$). The intermediate layer, undoped GaAs, acts as a tunnelling barrier. A 10 nm GaAs layer caps the quantum dots, and an AlAs/GaAs superlattice completes the structure in order to prevent current flow.

Our device, previously used for the control of dark exciton spin dynamics [18], is based on miniaturizing the active area of the charge-tunable structure to reduce the RC time constant. Photolithography was used to construct the miniaturized devices (Fig.1) out of low quantum dot density (<10 dots μm$^{-2}$) wafer sections, as detailed in [18]. A U-shaped Ohmic contact to the n-type layer was formed by annealing a layer of AuGe, Ni, and AuGe (60/10/60 nm respectively) deposited onto a section of the wafer surface. A 5 nm thick semi-transparent NiCr Schottky gate, of area less than 700 μm$^2$, was positioned a few microns from the centre of the U-shaped Ohmic contact layer. To minimize the capacitance, the back contact between the two arms of the Ohmic contact was removed by etching. A 360 nm thick NiCr layer was deposited onto the etched surface, making contact at one end with the Schottky gate. By removing the redundant back contact, this arrangement minimizes the stray capacitance.

Our device geometry serves two purposes. First, the Ohmic contact and contact strip form a coplanar waveguide impendence matched to 50 Ω, maximizing the coupling efficiency to our signal generator and high speed coaxial cabling. Secondly, the large length scales of the coplanar waveguide allow the Schottky gate to be positioned underneath a 0.9 mm diameter Weierstrass solid immersion lens (SIL) without compromising device performance. The SIL, with a refractive index n=2.15, provides a near ten-fold increase in the collection efficiency from a single quantum dot [22].

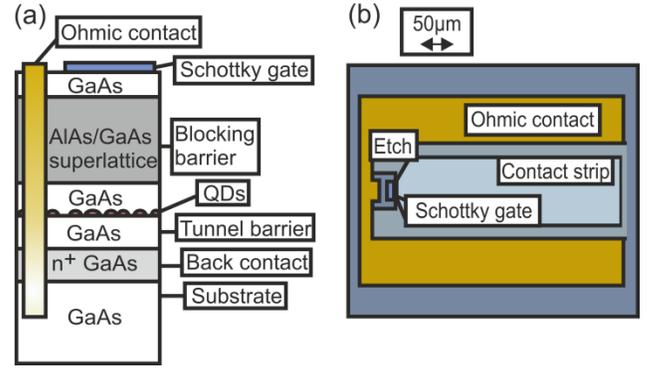

**FIG. 1.** Schematic view of the GHz bandwidth device. (a) The layer structure with the active region of quantum dots embedded between the tunnel barrier to the highly n-doped back contact and the capping layer. On top of the capping layer the AlAs/GaAs superlattice prevents current flow to the 5 nm thick, NiCr, micron-sized Schottky gate. The Ohmic contact between the back contact and the surface is made by annealing AuGe/Ni/AuGe layers. (b) Top view of the device, showing the 50 Ohm coplanar waveguide structure, the 400 nm deep etch and the contact strip for the miniaturized Schottky gate.

The device is connected to high speed, 50 Ω, brass SMA cabling using silver conductive paint. Voltage pulses are provided by an Agilent 81133A pulse pattern generator (PPG). The PPG generates voltage pulses up to 2 V with 60 ps 10-90% rise time. The voltage pulses propagate through the brass 50 Ω impedance coaxial cabling with less than 100 ps 10-90% rise time (measured at T=300 K). All experiments are carried out in a liquid helium cryostat at 4.2 K.

Single quantum dot optical excitation and photoluminescence (PL) collection is performed using confocal microscopy. The combination of Weierstrass SIL and 0.4 NA objective produces a collection spot size of ~0.25 μm$^2$. The device is mounted on a set of piezoelectric nanopositioners in order to scan the sample relative to the focus. Depending on the particular experiment, optical excitation is performed using either a non-resonant 830 nm continuous wave (cw) laser which excites carriers into the wetting layer, or a 1 MHz spectral bandwidth, tunable, cw external-cavity-diode laser which excites the optical transition resonantly. PL from the quantum dot is spectrally dispersed by a blazed grating spectrometer and detected using a liquid nitrogen-cooled charge-coupled device (CCD) camera, with a spectral resolution of ~50 μeV. A movable mirror within the spectrometer can be used to direct the PL to a secondary exit port where a ~0.5 meV spectral bandwidth section of the PL is collected by a 50 μm-core size multimode optical fibre, delivering the PL to a silicon single photon avalanche detector (SPAD). The SPAP has a full width at half maximum jitter of ~400 ps and is used to perform time-correlated single photon counting (TCSPC) of the PL. The detection efficiency of the SPAD is constant over the spectral range of these measurements.

The charge-tunable device allows the energy levels of the quantum dot to be manipulated precisely with respect to the Fermi level of the highly doped back contact via an applied voltage ($V_g$) between the back contact and the Schottky gate. Electrons tunnel into the quantum dot from the back contact as the quantum dot energy comes into resonance with the Fermi level. Each additional electron must overcome the strong Coulomb repulsion, giving rise to a pronounced Coulomb blockade, i.e. single electron charging, at progressively higher voltages [23]. Optical excitation provides holes, and the presence of a single hole reduces slightly the voltages of the Coulomb blockade plateau [24,25]. In this way it is possible to form to a single hole, the neutral exciton ($X^0$) and the negatively charged excitons $X^{1-}$, $X^{2-}$ and $X^{3-}$ which each contain 0, 1, 2, 3 and 4 electrons respectively, as a function of the applied bias. Under certain conditions the positively charged exciton, $X^{1+}$, can also form [26]. Single electron charging events are observed as discrete jumps in the PL energy due to the differing Coulomb energies for each exciton configuration. Fig. 2(a) shows an example plot PL as a function of $V_g$.

The miniaturized charge-tunable device has been designed to have a small capacitance and therefore small time response to an applied voltage pulse. Under ideal conditions, GHz bandwidth modulation would allow for the selection of exciton charge at a rate higher than the radiative recombination rate (~1 GHz) [27]. The intrinsic dynamics are fast as the electron tunnelling rate between the quantum dot and the back contact is ~100 GHz [28]. The challenge is to characterize the temporal response experienced by a single dot to an external driving voltage pulse. In this paper we utilize the PL signal from the quantum dot as a probe, taking advantage of the inherently fast tunnelling rates, to provide clear and concise information from a single quantum dot, under known conditions. The difficulty in translating this information to the opto-electronic characteristics for non-ideal devices is related to the equivalence of the device bandwidth and the radiative recombination rate. We present three separate, complimentary methods, of varying complexity, to extract directly the temporal voltage response from a single quantum dot.

### III. Switching of the charge state: the response to a voltage step

The exciton charge determines the energy at which PL emission occurs. Hence, time-dependent spectroscopy can yield information about a time-varying charge. With non-resonant excitation, the quantum dot is populated within several ps of excitation [28] with the excitonic configuration depending on the applied voltage $V_g$, as shown in Fig. 2(a). For instance, PL from the negatively charged exciton ($X^{1-}$) is only observed in the corresponding voltage range between 10 mV and 140 mV. In this first characterisation method the voltage plateau of $X^{1-}$ is used to probe the response to the GHz voltage pulse. The voltage pulse is applied from a voltage in the $X^0$ plateau to a voltage in the $X^{2-}$ plateau. Using the SPAD configuration, the experiment collects only emission from the $X^{1-}$ exciton. Whenever the voltage moves into the $X^{1-}$ plateau, a PL signal of the $X^{1-}$ exciton is observed. The time dependence of the $X^{1-}$ PL emission relative to the driving voltage provides a measure of the response of the quantum dot to the voltage pulse.

A single quantum dot was excited non-resonantly with 270 nW μm$^{-2}$ of 830 nm cw laser light. Figure 2(a) shows the PL emitted by the quantum dot against the applied voltage ($V_g$). The exciton configurations responsible for each PL line are labelled. $X^{1-}$ has a voltage extent of ~130 mV, from 10 mV< $V_g$ <140 mV. Two voltage points (marked $V_L$=-10 mV and $V_H$=160 mV in Fig. 2(a)) were chosen, both ~20 mV beyond the edge of the $X^{1-}$ plateau. The PPG was used to apply a 10 MHz square wave pulse with a rise time of 60 ps as a GHz bandwidth time-varying voltage between $V_L$ and $V_H$ (Fig. 2(b)). TCSPC was carried out, using a start voltage pulse synched to the PPG signal and a stop pulse triggered by the quantum dot PL when collected by the SPAD. Time resolved dynamics of $X^0$, $X^{1-}$ and $X^{2-}$ are shown in Fig. 2(c) (black dotted, red solid, and blue dashed lines respectively).

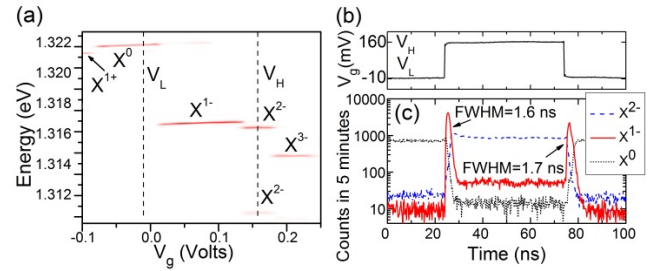

**FIG. 2.** The $X^{1-}$ voltage plateau under cw non resonant optical excitation as a probe of the response time of the device. (a) The PL for a range of $V_g$ for a single dot illuminated with 270 nW μm$^{-2}$ of 830 nm cw laser light. The scale depends linearly on the counts (the readout signal from the CCD camera), starting from white (less than 300 counts) with increasing intensity to red ≅10,000 counts. The exciton responsible for each PL line is identified. $V_L$=-10 mV and $V_H$=160 mV (dotted lines) are voltage points chosen to be 20 mV beyond either edge of the $X^{1-}$ voltage plateau. (b) An oscilloscope trace of the output of the PPG, showing a 10 MHz repetition rate square wave voltage pulse applied to the device between $V_L$ and $V_H$, as a function of time. (c) TCSPC measurements of $X^0$ (black dotted line), $X^{1-}$ (red solid line) and $X^{2-}$ (blue dashed line) with the square wave applied to the device.

When $V_g$=$V_L$ the device is at a bias within the $X^0$ voltage plateau. Consequently TCSPC on $X^0$ shows a large uncorrelated count rate, and the TCSPC of $X^{1-}$ and $X^{2-}$ show only background signal (arising from the SPAD dark counts). When $V_g$=$V_H$ the device is at a bias within the $X^{2-}$ voltage plateau. The TCSPC of $X^{2-}$ has a large uncorrelated count rate, the TCSPC of $X^0$ shows only background counts and the TCSPC of $X^{1-}$ shows a count rate which is slightly higher than the background count rate. This increased count rate from $X^{1-}$ is due to the detection of small amounts of $X^{2-}$ PL when centred on the $X^{1-}$ wavelength on account of the imperfect spectral filtering of the PL. When $V_g$ changes from $V_L$ to $V_H$ (at t=23 ns), the TCSPC signal of $X^0$ shows a rapid decrease in count rate, the TCSPC of $X^{2-}$ shows a rapid increase in count rate, and the TCSPC of $X^{1-}$ shows a peak in counts. Conversely, when $V_g$ changes from $V_H$ to $V_L$ (at t=73 ns) the TCSPC of $X^{2-}$ shows a rapid decrease in count rate, the TCSPC of $X^0$ shows a rapid increase in count rate, and the TCSPC of $X^{1-}$ shows a second peak in counts.

The peaks recorded at t=23 ns and t=73 ns in TCSPC of $X^{1-}$ show that PL is emitted from $X^{1-}$ during the voltage transition from $V_L$ to $V_H$. The full width at half maximum (FWHM) of each peak is ~1.6 ± 0.1 ns. This time can be considered as the total time taken for the voltage experienced by the quantum dot to pass through the $X^{1-}$ plateau. Taking the rise/fall time of the voltage applied by the PPG (~60 ps) into account, and the 400 ps response time of the SPAD, the voltage response time of the quantum dot is therefore approximately 1.5 ± 0.1 ns. We see similar results from other excitons in the same quantum dot and from other quantum dots in the same sample.

This method is simple to perform and gives an approximate measure of a crucial time in applications, the time taken to traverse a Coulomb blockade plateau. However, it lacks the ability to map the specific voltage response, which is a potentially limiting factor at higher repetition rates. Nevertheless, these results demonstrate near GHz response behavior from a single quantum dot under external electrical modulation.

## IV. Switching of the charge state: the response to a voltage pulse

Our second method for measuring the voltage response time of the quantum dot follows a more complex approach but provides, in addition to the response time, an indirect measurement of the temporal form of the response. This approach is based on the attenuation of a voltage pulse when applied to a device with smaller response time than the pulse width. If the pulse duration (ΔT) of an input voltage pulse is much larger than the response time (τ) of the device, there is no pulse attenuation. On the other hand, if the pulse duration is comparable to or shorter than τ, there will be an inherent pulse attenuation as the device is unable to respond to the full dynamic range of the input signal. Here, a measurement of the pulse attenuation against the pulse duration produces a value for τ, as well as the temporal form of the quantum dot voltage response. Consequently, unlike Methods 1 and 3, the PL spectra are recorded under steady state conditions, without the need for TCSPC. This was performed on the same quantum dot as the method described in section III.

The quantum dot was illuminated exactly as in Method 1, with 270 nW μm$^{-2}$ of 830 nm cw laser light. Figure 3(a) shows again the PL emitted by the quantum dot against $V_g$, but this time with different labelling specific to this measurement method. A square wave voltage pulse (rise/fall time < 100 ps, repetition rate = 20 MHz) was applied to the device by the PPG, Fig. 3(b). The voltage amplitude, $V_H - V_L$, was fixed at 200 mV.

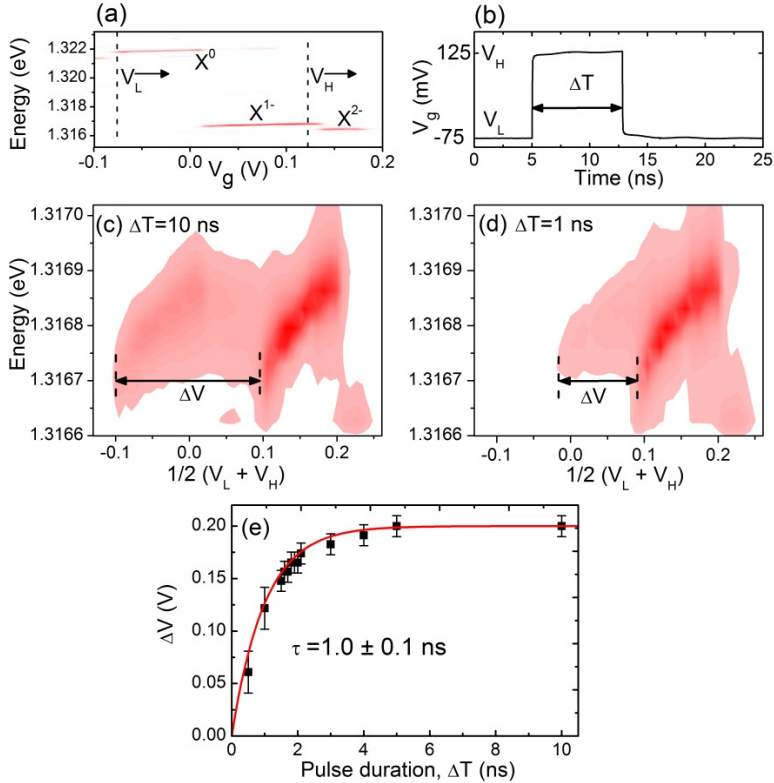

**FIG. 3.** The signal truncation of a voltage pulse applied to the quantum dot device. (a) PL against $V_g$, for the same dot as Fig 2(a), but now shown for a smaller range. (b) An oscilloscope trace of the 20 MHz repetition rate voltage pulse applied to the device between $V_L$ and $V_H$ as a function of time. The voltage difference between $V_L$ and $V_H$ is kept constant at 200 mV. The voltage offset (1/2 ($V_L + V_H$)) and the pulse duration (ΔT) were varied, indicated by the arrows in (a) and (b). (c) The $X^{1-}$ PL against voltage offset for the dot from (a), with a time-varying applied voltage as in (b), for ΔT=10 ns. There are two PL lines due to emission from $X^{1-}$ at different offset voltages. The voltage difference between the onset of each PL line is marked as ΔV. (d) The PL against voltage offset for the dot from (a), with a time-varying applied voltage as in (b) but with ΔT=1 ns. (e) ΔV measured for various values of ΔT (black squares), alongside an exponential fit (ΔV = ΔV$_0$ (1- exp(-t/τ))) with an amplitude of ΔV$_0$=200 mV, and a 1/e rise time of τ=1 ns. All data shown were taken with 270 nW μm$^{-2}$ of 830 nm CW laser light.

Whilst keeping the pulse duration (ΔT) fixed the $X^{1-}$ PL spectra were recorded as the voltage offset (the mean value of $V_H$ and $V_L$) was varied. The experiment was then repeated for different values of ΔT. Figure 3(c) shows the results for ΔT =10 ns, where PL from $X^{1-}$ is observed at two separate voltage offset regions. At lower voltage offsets (-0.1 V to 0.02 V) $X^{1-}$ PL is recorded because VH lies within the $X^{1-}$ voltage plateau, whereas at higher voltage offsets (0.1 V to 0.23 V) $X^{1-}$ PL is recorded because VL lies within the $X^{1-}$ voltage plateau. The voltage difference between the onset of the $X^{1-}$ PL extent at lower and higher offsets (ΔV) was 200 mV, signifying that for ΔT = 10 ns there is no notable attenuation of the driving voltage pulse.

Figure 3(d) shows the PL from $X^{1-}$ under the same conditions as Fig. 3(c) only with ΔT = 1 ns. For this pulse duration ΔV is reduced to ~120 mV. This is because the quantum dot takes a finite time to experience a change from VL to VH. As ΔT becomes comparable to this response time, the voltage experienced by the quantum dot no longer has the time to reach VH. As such, ΔV measures the voltage experienced by the quantum dot at time ΔT.

The measured value of ΔV relative to various values of ΔT is shown in Fig. 3(e). As ΔT becomes smaller, ΔV decreases. The data points are fitted well by a single exponential with a 1/e rise time (τ) of 1.0 ± 0.1 ns. Although of comparable magnitude, this value is slightly smaller than the 1.5 ns determined with Method 1 for the same quantum dot. Even though this approach successfully maps the voltage response from a single quantum dot, measurements of ΔV are difficult to perform when ΔT is small (at low PL rates), hence there is a larger error in the measurements of ΔV at small values of ΔT.

**V. The hybridization region as a probe of the response to a voltage step**

An alternative approach to the first method described, where the voltage extent used as the probe is similar in magnitude to the applied pulse amplitude, is to use a probe region which is significantly smaller in extent than the voltage pulse. Such an approach would increase the precision in mapping the voltage response and reduces complexities related to internal exciton dynamics. However, the implementation is somewhat challenging. The simplest approach would be to increase the voltage amplitude to several volts, around 10 V, and to use the exciton emission over the extent of the entire charging regime, ~0.5 V, as the probe. However this has several limitations. First, selecting only one exciton is unrealistic if the voltage amplitude is very large as the quantum dot emission can become spectrally very broad once the wetting layer is occupied at higher gate voltage. Secondly, large voltages, beyond a few volts, can lead to breakdown of a Schottky diode device. Finally, such voltages are unrealistic for real world applications and are impossible for many GHz sources, our PPG included. Our alternative approach is to exploit the "hybridization region" **H**, a narrow region at the low bias end of the $X^{1-}$ plateau, in which resonant excitation of $X^0$ leads to $X^{1-}$ emission [29,30]. In this case the probe region is much smaller than our 200 mV applied voltage pulse.

The PL against $V_g$ is recorded in Fig. 4(a), for a single quantum dot excited by 5.5 nW μm$^{-2}$ of non-resonant cw 830 nm laser light, with the exciton emission labelled. **H** is the hybridization region in which the lowest stable state with one hole in the quantum dot is $X^{1-}$, and the lowest stable state without a hole is an empty quantum dot (|0>) [29,31]. For a given $V_g$ within **H**, the resonant excitation at 1.3023 eV populates the quantum dot with an $X^0$ exciton. An electron tunnels into the quantum dot from the back contact, forming $X^{1-}$, on a time scale (~50 ps) much faster than the radiative recombination time of the exciton (~1 ns). After the $X^{1-}$ decay via electron-hole recombination, a single electron is left in the quantum dot, which then tunnels to the back contact leaving the quantum dot empty. The system is therefore reset, and the quantum dot can be re-excited by the laser. Figure 4(c) shows the result of this cycle. The quantum dot is excited with 73.4 μW μm$^{-2}$ of resonant cw laser light and tuned to 1.3023 eV. $X^{1-}$ PL is observed over a voltage region of ~24 mV. The small amount of laser light also seen in the contour is due to limitations of the filtering method used before imaging with the CCD. However, this has no adverse effect on the SPAD measurement as the SPAD collection geometry provides a second level of spectral filtering that effectively removes the laser signal. The voltage region over which the $X^{1-}$ PL emits in Fig. 4(c) is shifted relative to the region **H** marked in Fig. 4(a), due to a reduced degree of hole storage in the capping layer with resonant as opposed to non-resonant excitation [32, 33].

A 10 MHz square wave time-varying voltage, with 50/50 duty cycle (shown in Fig. 4(b)), was applied to the sample with a fixed amplitude of 200 mV and with a tunable voltage offset as described in Method 2. We use here a different quantum dot from Methods 1 and 2, embedded within a separate device. However, both devices have been manufactured under similar conditions, to the same specifications and from the same wafer material. Figure 4(d) is a TCSPC measurement of $X^{1-}$, showing the interaction of the quantum dot with the laser as the voltage changes from $V_H$ = 275 mV to $V_L$ = 75 mV at 26 ns and the opposing voltage rise from 75 mV to 275 mV at 76 ns. Two peaks are observed which can be attributed to the points at which $V_g$ enters or leaves the voltage region **H**. Two points on each PL peak are defined: **A** is at the maximum of each peak; **B** is at half of maximum on the rising edge of each peak. The time of occurrence of **A** and **B** were recorded as a function of the voltage offset (Fig. 4(e) and Fig. 4(f), respectively).

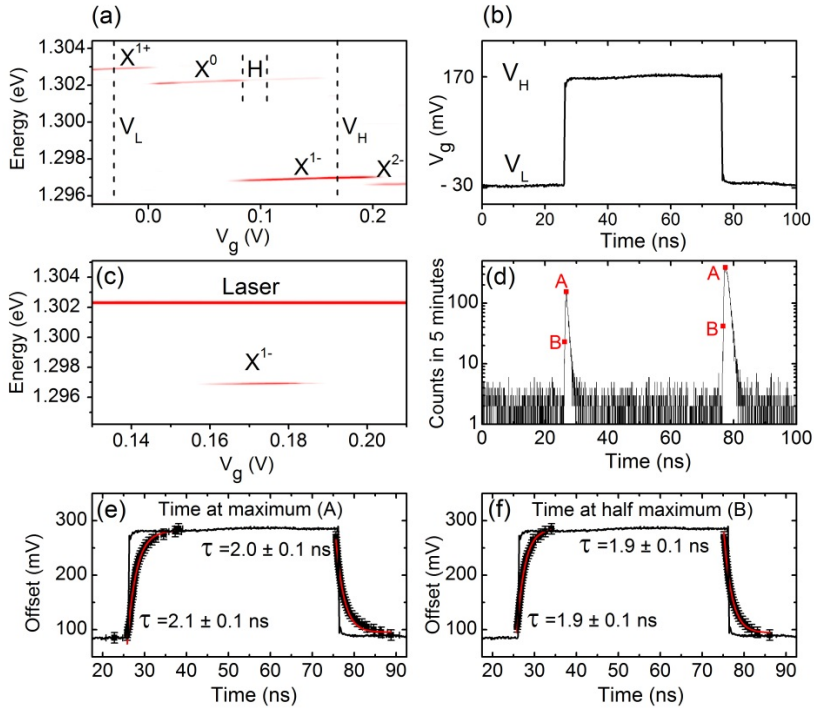

**FIG. 4.** TCSPC of $X^{1-}$ following resonant excitation of $X^0$. (a) PL against $V_g$ for a dot illuminated with 5.5 nW μm$^{-2}$ of non-resonant 830 nm cw laser light. The exciton responsible for each PL line is identified. The region **H** marks the hybridization region, a voltage extent in which resonant excitation of $X^0$ results in emission of $X^{1-}$ PL. (b) An oscilloscope trace of the 10 MHz repetition rate square wave, with 200 mV amplitude. (c) Spectra for the dot in (a) as a function of $V_g$ when the dot is illuminated with 73.4 μW μm$^{-2}$ of laser light tuned to 1.3023 eV, resonant with the $X^0$ transition of the dot. PL emission from $X^{1-}$ marks the voltage region **H**. (d) TCSPC of $X^{1-}$, for the dot from (a), illuminated with 73.4 μW μm$^{-2}$ of light tuned to 1.3023 eV, with a time-varying applied voltage as in (b), with an offset of 175 mV. Two PL peaks are observed at 26 ns and 76 ns, and a maximum (**A**) and half maximum (**B**) point on each peak are identified. (e) The time of occurrence of peak **A** as a function of the voltage offset ($1/2$ ($V_L + V_H$)) with the applied pulse from (b). (f) The time of occurrence of **B** as a function of the voltage offset with the applied pulse from (b). (e) and (f) show exponential fits to the rise/fall.

The data provide an accurate picture of the quantum dot response to an applied voltage. An exponential function fits the data very well with a value of ~1.9 ± 0.2 ns for the 1/e voltage response time of the quantum dot, from both measurement points A and B. This value is larger than found for the previous device. This is most likely due to small fabrication differences, for example the size of the Schottky gate, the resistivity of the contact layers or the contacting to the sample. This method is more difficult to employ due to the use of resonant excitation. It provides however a very precise idea of the temporal form of the quantum dot voltage response.

## VI. Conclusions

Three separate methods, all based on exploiting the optical response from well-defined exciton states, have been developed to measure the voltage response in a GHz bandwidth of a single InGaAs quantum dot embedded within a microstructured charge-tunable device at low temperature to a fast voltage pulse. The form of the quantum dot response to an applied voltage was found to be a single exponential using two separate techniques performed on different quantum dots within different devices. Each method showed consistent results. In the best case, a value of 1.0 ± 0.1 ns was obtained for the response time of a single quantum dot in these devices, confirming that GHz electrical control has been achieved. These results pave the way to the application of GHz electronics to the opto-electronic control of excitons and spins in self-assembled quantum dots.

We acknowledge financial support from the EPSRC (UK), the Royal Society University Research Fellowship, the Swiss National Science Foundation (SNF) and NCCR QSIT.

[1] J. Q. You and F. Nori, Nature **474,** 589 (2011).
[2] R. J. Schoelkopf and S. M. Girvin, Nature **451,** 664 (2008).
[3] J. R. Petta, A. C. Johnson, J. M. Taylor, E. A. Laird, A. Yacoby, M. D. Lukin, C. M. Marcus, M. P. Hanson, and A. C. Gossard, Science **309,** 2180 (2005).
[4] K. C. Nowack, F. H. L. Koppens, Yu. V. Nazarov, and L. M. K. Vandersypen, Science **318,** 1430 (2007).
[5] C. Ospelkaus, U. Warring, Y. Colombe, K. R. Brown, J. M. Amini, D. Leibfried, and D. J. Wineland, Nature **181,** 476 (2011).
[6] A. Shields, Nature Photonics **1,** 215 (2007)
[7] M. Atatüre, J. Dreiser, A. Badolato, A. Högele, K. Karrai, and A. Imamoglu, Science **312**, 551 (2008).
[8] D. Press, T. D. Ladd, B. Zhang, and Y. Yamamoto, Nature **456,** 218 (2008).
[9] R.-B. Liu, W. Yao, and L.J. Sham, Advances in Physics **59,** 703 (2010).
[10] B. D. Geradot, D. Brunner, P. A. Dalgarno, P. Öhberg, S. Seidl, M. Kroner, K. Karrai, N. G. Stoltz, P. Petroff, and R. J. Warburton, Nature **451**, 441 (2008).
[11] D. Brunner, B. D. Geradot, P. A. Dalgarno, G. Wüst, K. Karrai, N. G. Stoltz, P. M. Petroff, and R. J. Warburton, Science **325**, 70 (2009).
[12] K. De Greve, P. L. McMahon, D. Press, T. D. Ladd, D. Bisping, C. Schneider, M. Kamp, L. Worschech, S. Höfling, A.

Forchel, and Y. Yamamoto, Nature Physics, Advanced Online Publication (2011).
[13] A. Greilich, S. G. Carter, K. Kim, A. S. Bracker, and D. Gammon, Nature Photonics, Advanced Online Publication (2011).
[14] H. Drexler, D. Leonard, W. Hansen, J. P. Kotthaus, and P. M. Petroff, Phys. Rev. Lett. **73,** 2252 (1994).
[15] R. J. Warburton, C. Schäflein, D. Haft, F. Bickel, A. Lorke, K. Karrai, J. M. Garcia, W. Schoenfeld, and P. M. Petroff, Nature **405,** 926 (2000).
[16] K. J. Vahala, Nature **424,** 839 (2003)
[17] J. Claudon, J. Bleuse, N. S. Malik, M. Bazin, P. Jaffrennou, N. Gregersen, C. Sauvan, P. Lalanne, and J-M. Gérard, Nature Photonics **4,** 174 (2010)
[18] J. McFarlane, P. A. Dalgarno, B. D. Gerardot, R. H. Hadfield, R. J. Warburton, K. Karrai, A. Badolato, and P. M. Petroff, App. Phys. Lett. **94,** 093113 (2009).
[19] S. M. de Vasconcellos, S. Gordon, M. Bichler, T. Meier, and A. Zrenner, Nature Photonics **4,** 545(2010).
[20] D. Kim, S. G. Carter, A. Greilich, A.S. Bracker, and D.Gammon, Nature Physics **7,** 223 (2011).
[21] R. H. Hadfield, Nature Photonics **3,** 696 (2009)
[22] K. A. Serrels, E. Ramsey, P. A. Dalgarno, B. D. Gerardot, J. A. O'Conner, R. H. Hadfield, R. J. Warburton, and D. T. Reid, Journal of Nanophotonics **2,** 021854 (2008).
[23] B. T. Miller, W. Hansen, S. Manus, R. J. Luyken, A. Lorke, J. P. Kotthaus, S. Huant, G. Medeiros-Ribeiro, and P. M. Petroff, Phys. Rev. B **56,** 6764 (1997).
[24] S. Seidl, M. Kroner, P. A. Dalgarno, A. Hogele, J. M. Smith, M. Ediger, B. D. Gerardot, J. M. Garcia, P. M. Petroff, K. Karrai, and R. J. Warburton, Phys. Rev. B **72,** 195339 (2005).
[25] S. Laurent, B. Eble, O. Krebs, A. Lemaître, B. Urbaszek, X. Marie, T. Amand, and P. Voisin, Phys. Rev. Lett. **94,** 147401 (2005).
[26] M. Ediger, P. A. Dalgarno, J. M. Smith, B. D. Gerardot, K. Karrai, P. M. Petroff, and R. J. Warburton, Appl. Phys. Lett. **86,** 211909 (2005).
[27] P. A. Dalgarno, J. M. Smith, J. McFarlane, B. D. Gerardot, K. Karrai, A. Badolato, P. M. Petroff, and R. J. Warburton, Phys. Rev. B **77,** 245311 (2008).
[28] J. M. Smith, P. A. Dalgarno, R. J. Warburton, A. O. Govorov, K. Karrai, B. D. Gerardot, and P. M. Petroff, Phys. Rev. Lett. **94,** 197402 (2005).
[29] C. Kloeffel, P. A. Dalgarno, B. Urbaszek, B. D. Geradot, D. Brunner, P. M. Petroff, D. Loss, and R. J. Warburton, Phys. Rev. Lett. **106,** 046802 (2011).
[30] C.-M. Simon, T. Belhadj, B. Chatel, T. Amand, P. Renucci, A. Lemaître, O. Krebs, P. A. Dalgarno, R. J. Warburton, X. Marie, and B. Urbaszek, Phys. Rev. Lett. **106,** 166801 (2011).
[31] P. A. Dalgarno, M. Ediger, B. D. Gerardot, J. M. Smith, S. Seidl, M. Kroner, K. Karrai, P. M. Petroff, A. O. Govorov, and R. J. Warburton, Phys. Rev. Lett. **100,** 176801 (2008).
[32] J. M. Smith, P. A. Dalgarno, B. Urbaszek, E. J. McGhee, G. S. Buller, G. J. Nott, R. J. Warburton, J. M. Garcia, W. Schoenfeld, and P. M. Petroff, App. Phys. Lett. **82,** 3761 (2003).
[33] B. Urbaszek, E. J. McGhee, J. M. Smith, R. J. Warburton, K. Karrai, B. D. Gerardot, J. M. Garcia, and P. M. Petroff, Physica E **17,** 35 (2003).